\documentclass[prd,aps,preprint,amsmath,nofootinbib,amssymb,eqsecnum,showkeys]{revtex4-1}
\pdfoutput=1
\usepackage{verbatim,graphics,graphicx,color,slashed}
\usepackage{ulem} 
\usepackage[colorlinks=true,
            linkcolor=red,
            urlcolor=blue,
            citecolor=blue]{hyperref}

\graphicspath{{figures/}}

\newcommand{\eV}{\mathrm{eV}}

\newcommand{\MeV}{\mathrm{MeV}}

\begin{document}
\title{More Is Different: Reconciling eV Sterile Neutrinos with Cosmological Mass Bounds}
\author{Yong Tang}
\affiliation{School of Physics, Korea Institute for Advanced Study,
 Seoul, South Korea}
\date{\today}

\begin{abstract}
It is generally expected that adding light sterile species would increase the effective number of neutrinos, $N_{\textrm{eff}}$. In this paper we discuss a scenario that $N_{\textrm{eff}}$ can actually decrease due to the neutrino oscillation effect if sterile neutrinos have self-interactions. We specifically focus on the eV mass range, as suggested by the neutrino anomalies. With large self-interactions, sterile neutrinos are not fully thermalized in the early Universe because of the suppressed effective mixing angle or matter effect. As the Universe cools down, flavor equilibrium between active and sterile species can be reached after big bang nucleosynthesis (BBN) epoch, but leading to a decrease of $N_{\textrm{eff}}$. In such a scenario, we also show that the conflict with cosmological mass bounds on the additional sterile neutrinos can be relaxed further when more light species are introduced. To be consistent with the latest {\textit{Planck}} results, at least 3 sterile species are needed.  
\end{abstract}

\maketitle

\section{Introduction}
Although most neutrino experiments can be well described by the standard three light species paradigm, there have been several anomalies that indicate a new sterile state~\cite{LSND:1996jb,LSND:2001ty,Reactor:2011rk,Flux:2005tb,MiniBooNE:2010wv,MiniBooNE:2013pmq}. The new state should have an eV scale mass and a mixing angle with $\sin^2 2\theta_0\sim 0.01$~\cite{Kopp:2013vaa,Giunti:2013aea}. With such parameters, sterile neutrinos can be copiously produced from oscillation with each species increasing the effective number of neutrinos $N_{\textrm{eff}}$ almost by one unit. This would be in tension with the cosmological bounds from cosmic microwave background (CMB) data~\cite{Planck:2013zuv}, $N_{\textrm{eff}}<3.91$ and $m^{\textrm{eff}}_{\nu}<0.59 \textrm{eV}$.

The conflict can be resolved if the sterile species is only partially thermalized with $N_{\textrm{eff}}\ll 1$. Partial thermalization can be realized in particle physics models where there are secret self-interactions\footnote{In this paper, we do not discuss the case that large lepton asymmetry exists in the active neutrino sector.} in the sterile neutrino sector~\cite{Hannestad:2013ana,Kopp:2013zpn,Bringmann:2013vra,Ko:2014bka,Hannestat:2014nda}. These self-interactions can induce large matter potentials, effectively suppress the mixing angle, and block sterile neutrino's production from oscillation efficiently~\cite{Hannestad:2013ana,Kopp:2013zpn}. 

However, the situation changed recently when authors in~\cite{Mirizzi:2014ama} argued that even if sterile neutrinos' production are blocked at BBN time, as the Universe cools down, the new self-interaction will eventually equilibrate sterile neutrinos with the active ones. Then, it can be easily shown that in 3+1 scenario, $1/4$ of the cosmic background neutrinos would be the heavy sterile ones, which is still in conflict with the above cosmological bounds.

The novelty in this paper is that, we numerically solve the quantum kinetic equations for neutrino mixing, show and confirm that flavor equilibrium is indeed reached after BBN, see Fig.~\ref{fig:neffGx}. We also propose that the above mentioned conflict can be reconciled easily in an extension that more than one sterile states are introduced. Of the $n$ introduced self-interacting sterile species, only one has eV scale mass and mixes with the active neutrinos. In the early Universe, all of them are out of equilibrium, but can approach flavor equilibrium with active neutrinos after BBN era. So the current relic neutrinos are composed of 3+n species with equal flavors. As we shall show that $N_{\textrm{eff}}$ in the late Universe can actually decrease and the cosmological bounds can be evaded. To be consistent with the latest {\textit{Planck}} results, we found that $n=3$ is the minimal number of the introduced sterile species.

This paper is organized as follows. In section~\ref{sec:overview}, we briefly review the effective number of neutrinos, $N_{\textrm{eff}}$.  In section~\ref{sec:neff}, we discuss the scenario that how $N_{\textrm{eff}}$ in BBN epoch can be different from its value in CMB time when there are  secret self-interactions among sterile neutrinos. In section~\ref{sec:massbounds}, we show how the cosmological mass bounds on sterile neutrino can be relaxed if more than one light species are introduced. Finally, we give our conclusion.

\section{Overview of $N_{\mathrm{eff}}$}\label{sec:overview}
In this section, we shall review $N_\textrm{eff}$ briefly and establish the related conventions and definitions.
 
First, let us recall the thermal history of neutrinos in the standard model (SM) at the early Universe. When the temperature of the thermal bath, $T_{\gamma}$, is much higher than MeV, active neutrinos, $\nu_a (a=e,\mu,\tau)$, are in thermal equilibrium with other SM particles through electroweak processes and have the same temperature as $T_{\gamma}$. Around 2 MeV, active neutrinos are decoupled because electroweak interaction is not strong enough to keep them in equilibrium. Later, electrons/positrons annihilate but heat only the photons. Using the conservation of entropy density, one can obtain the temperature ratio after $e^{\pm}$ annihilation,
\begin{equation}\label{eq:tempratio}
\frac{T_{\nu_a}}{T_\gamma}=\left(\frac{4}{11}\right)^{1/3}.
\end{equation}
Afterwards, $T_{\nu_a}/T_\gamma$ is constant in the standard cosmology.

$N_\textrm{eff}$ is defined by the energy density ratio,
\begin{equation}\label{eq:neff}
\frac{\rho_R}{\rho_\gamma}=\frac{\rho_\gamma+\sum_a\rho_{\nu_a}}{\rho_\gamma}=1 + \frac{7}{8}\left(\frac{T_{\nu_a}}{T_\gamma}\right)^4 N_\textrm{eff},
\end{equation}
where $\rho_R$ stands for the total energy density of radiations. In the standard model, we have 3 species of active neutrinos, $\nu_e,\nu_{\mu}$ and $\nu_{\tau}$, so $N_{\textrm{eff}}=3$, or $3.046$ precisely if instantaeneous neutrino decoupling is relaxed.
If there is some new physics, it might modify $T_{\nu_a}/T_\gamma$ and/or contribute to $\rho_{R}$ as extra radiation. For example, dark matter may affect $T_{\nu_a}$~\cite{Boehm:2012gr, Steigman:2014lwa} or $T_\gamma$~\cite{Ho:2012br}, and new particles can decay to or contribute as equivalent neutrinos~\cite{Giunti:2014pja,Hasenkamp:2014hma,Huang:2014dpa}. In this paper, we only consider the extended models with sterile neutrinos. 

In the following discussions, we shall use $\nu_a (a=e,\mu,\tau)$ to denote the 3 active neutrinos, $\nu_s$ for the sterile species, and $\nu_\beta$ or $\nu$ for all of them if not otherwise specified. And $\nu_i$ will be referred as the $i$-th mass eigenstate with a mass $m_{\nu_i}$. For the parameter space we focus on,  $\nu_{1,2,3}$ are mainly mixed states of active neutrinos $\nu_a$ and $\nu_{i}$s $(i>3)$ are mainly mixed sterile species. We shall also neglect the masses of $\nu_{1,2,3}$ when considering the mass constraints on sterile neutrinos. 

In principle, $N_{\textrm{eff}}$ can be a function of time or photon temperature. We define $\delta N_{\textrm{eff}}$ as the deviation from the standard value, with explicit time/temperature dependence,
\begin{equation}\label{eq:Neff}
\frac{\rho_R-\rho_\gamma}{\rho_\gamma}=\frac{7}{8}\left(\frac{T_{\nu_a}^0}{T_\gamma}\right)^{4}
\left[3+\delta N_{\textrm{eff}}\left(t \right)\right],
\end{equation}
or
\begin{equation}\label{eq:deltaNeff}
\delta N_{\textrm{eff}}\left(t\right)=\sum_{\beta} 
\left(\frac{T_{\nu_\beta}\left(t \right)}{T_{\nu_a}^0}\right)^4-3,
\end{equation}
where $T_{\nu_a}^0$ stands for the neutrino temperature in the standard cosmology without new physics, $T_{\nu_a}^0=T_\gamma$ before $e^{\pm}$ annihilation and $T_{\nu_a}^0=\left(4/11\right)^{1/3}T_\gamma$ afterwards,  and $\beta$ runs through all active/sterile neutrinos.

We shall keep in mind that $3$ in Eqs.~\ref{eq:Neff} and \ref{eq:deltaNeff} is actually $3.046$ precisely. But this little difference would not affect our later discussions and we shall use $3$ throughout the paper. 

\section{$N^{\textrm{bbn}}_{\mathrm{eff}}$ vs $N^{\textrm{cmb}}_{\mathrm{eff}}$} \label{sec:neff}

\begin{figure}[t]
\includegraphics[scale=0.8]{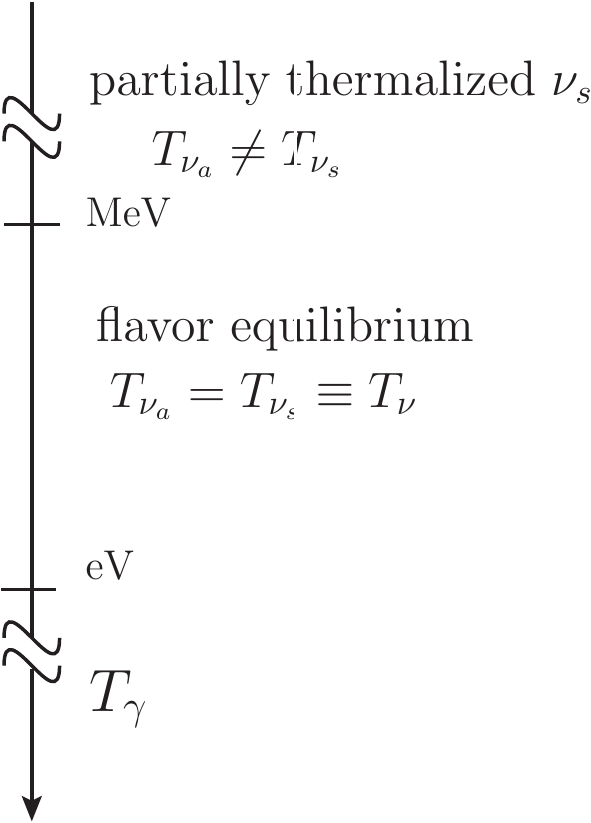}
\caption{Thermal history of active/sterile neutrinos. When the temperature is high, $\nu_s$s are not in thermal equilibrium with $\nu_a$s because of the suppression from a large matter potential. As the Universe cools down, equilibrium between active and sterile neutrinos could be reached.  \label{fig:thermalhistory}}
\end{figure}

In this section, we discuss how self-interaction can affect the thermalization of sterile neutrinos. The essential picture is described in Fig.~\ref{fig:thermalhistory} where sterile neutrinos are only partially thermalized at/before BBN time, but flavor equilibrium, $\rho_{\nu_s}=\rho_{\nu_a}$, is reached at later time. To be as general as possible, we introduce $n$ sterile species and do not discuss the specific particle physics models, but emphasize that new interaction for the sterile neutrinos is required.

It should be pointed out that in the present discussion for illustrating our points, the minimal setup we need is that {\it one} of the $n$ sterile species is mixed with active neutrinos and has eV-scale mass, and the rest may be massless and have negligible mixing to affect current neutrino experimental results. Moreover, sterile neutrinos do not have to mix with each other since all sterile neutrinos can be in flavor equilibrium through new interactions.
Therefore, let us just simply assume there is only mixing between a 4-th neutrino and the active species, namely the mixing matrix has the following form,
\begin{equation}
\left(
\begin{array}{c}
 \nu_{e} \\ \nu_{\mu} \\ \nu_{\tau} \\ \nu_s \\ \vdots
\end{array}
\right) = 
\left ( \begin{array}{ccccc} 
    U_{e1} 	  & U_{e2}   & U_{e3}   & U_{e4}    &  \cdots \\ 
	U_{\mu1}  & U_{\mu2} & U_{\mu3} & U_{\mu 4} &  \cdots \\
	U_{\tau1} & U_{\tau2}& U_{\tau3}& U_{\tau 4}&  \cdots \\
	U_{s1}    & U_{s2} 	 & U_{s3}   & U_{s4}    &  \cdots \\
    \vdots    & \vdots   & \vdots   & \vdots    & 1
\end{array} 
\right)
\left( 
\begin{array}{c}
\nu_{1} \\ \nu_{2} \\ \nu_{3} \\ \nu_{4} \\ \vdots 
\end{array}
\right).
\end{equation}
Further assume only $1$-$4$ mixing and CP conservation, then the complete $4\times 4$ mixing matrix is 
\begin{eqnarray}
&&U_{4\times 4}=
\left( 
\begin{array}{cccc}
 c_{13}c_{12} & c_{13}s_{12} & s_{13}  & 0\\
 -c_{23}s_{12}-s_{13}s_{23}c_{12} & 
c_{23}c_{12}-s_{13}s_{23}s_{12} &
c_{13}s_{23} & 0 \\
 s_{23}s_{12}-s_{13}c_{23}c_{12} & 
-s_{23}c_{12}-s_{13}c_{23}s_{12} &
c_{13}c_{23} & 0 \\
0 & 0 & 0& 1
\end{array} 
\right) \cdot
        \left(
         \begin{array}{cccc}
          c_{14}& 0& 0& s_{14}\\
          0& 1& 0& 0\\
          0& 0& 1& 0\\
          -s_{14}& 0&0& c_{14} \\
         \end{array}
        \right) \nonumber \\
&&=
\left( 
\begin{array}{cccc}
 c_{13}c_{12}c_{14} & c_{13}s_{12} & s_{13}  & c_{13}c_{12}s_{14}\\
 -c_{23}s_{12}c_{14}-s_{13}s_{23}c_{12}c_{14} & 
c_{23}c_{12}-s_{13}s_{23}s_{12} &
c_{13}s_{23} &  -c_{23}s_{12}s_{14}-s_{13}s_{23}c_{12}s_{14}\\
 s_{23}s_{12}c_{14}-s_{13}c_{23}c_{12}c_{14} & 
-s_{23}c_{12}-s_{13}c_{23}s_{12}e^{i\delta_{CP}} &
c_{13}c_{23} & s_{23}s_{12}s_{14}-s_{13}c_{23}c_{12}s_{14}\\
-s_{14} & 0 & 0& c_{14}
\end{array} 
\right).        \nonumber \\
\end{eqnarray}
where e.g. $c_{ij} = \cos\theta_{ij}$ and $s_{ij} = \sin\theta_{ij}$, etc.
Complete investigations with the above multiple-flavor mixing are quite involved numerically when solving the full quantum kinetic equations (QKEs)~\cite{Sigl:1992fn,Thomson:1992ja,Stodolsky:1986dx,Harris:1980zi} and we refer to Refs.~\cite{Melchiorri:2008gq,Hernandez:2013lza,Hernandez:2014fha} for multiple-flavor analysis with non-interacting sterile neutrinos. For simplicity and without loss of generality, we work with only two neutrino states, $\nu_e$-$\nu_s$ mixing with mass difference $\delta m^2$ and mixing angle $\theta_0\equiv \theta_{14}$, and shall pay our attention to eV sterile neutrinos with the parameter space suggested by neutrino anomalies, $\delta m^2 \simeq 1\textrm{eV}^2$ and $\sin^2 2\theta_0\simeq 0.01$. We shall keep in mind that $v_s$ produced from $\nu_\mu$ and $\nu_\tau$'s oscillation could be equally important since $|U_{\mu 4}|\simeq |U_{\tau 4}| \simeq 0.46 |U_{e 4}|$.

In most of the previous discussions in the literature, sterile neutrinos are assumed to have no interaction, so they can only be produced by oscillations from the active ones. And the production rate and total amount depend on the mass difference, mixing angle and lepton asymmetry~\cite{Foot:1996qc,Fuller:1998km,Fuller:2001nj,Chu:2006ua,Dodelson:1994,Dolgov:1990vx,Bell:1998ds,Enqvist:1990ad,Enqvist:1991qj,Sorri:2001cb,Krauss:2013xr,Khlopov:1981nq}.

In case of sterile species having a secret self-interaction, parametrized by $G_X\equiv g^2_X/M^2_X$ (similar to the Fermi's constant $G_F$ in SM), 
\begin{equation}
G_X \bar{\nu}_s\Gamma_i \nu_s\; \bar{\nu}_s\Gamma_j \nu_s,\;\Gamma_{i,j} \textrm{ are products of }\gamma_{\mu},\gamma_5,...,
\end{equation}
the production will also depend on the strength of $G_X$. Models that can give rise to the above types of self-interactions can be found in Refs. \cite{Hannestad:2013ana, Kopp:2013zpn, Bringmann:2013vra, Ko:2014bka, Hannestat:2014nda, Shoemaker:2014xra,Nelson:2014mva}. To calculate the $N^{\textrm{bbn}}_{\textrm{eff}}$, we use the modified version of \texttt{LASAGNA}~\cite{Hannestad:2013ana, Hannestad:2012ky, Hannestad:2013pha} with $g_X=0.1$.

For $1$+$1$ scenario, we can parametrize the system with $2\times 2$ Hermitian density matrix in terms of Pauli matrices,
\begin{equation}
\rho=\frac{1}{2}f_0\left(P_0 + \vec{P}\cdot \vec{\sigma}\right),
\end{equation}
where $f_0=1/\left(e^{E/T}+1\right)$ and $\vec{P}=\left(P_x,P_y,P_z\right)$. And the densities of active and sterile neutrino are given by
\begin{equation}
P_a\equiv P_0 + P_z=2\frac{\rho_{\nu_a}}{f_0}, P_s\equiv P_0-P_z=2\frac{\rho_{\nu_s}}{f_0}.
\end{equation}
The kinetic equations governing $P_i$'s time evolution are 
\begin{align}
\dot{P}_a &= V_x P_y + \Gamma_a \left[ 2 \frac{f_{eq,a}}{f_0}- P_a \right],  \\
\dot{P}_s &= -V_x P_y + \Gamma_s \left[ 2 \frac{f_{eq,s}}{f_0}- P_s \right], \\
\dot{P}_x &= -V_z P_y -D P_x, \\
\dot{P}_y &= V_z P_x -\frac{1}{2}V_x (P_a -P_s)-D P_y,
\end{align}
where $V_{i}$ and $\Gamma_i$ are the potentials and scattering kernels~\cite{Hannestad:2013ana}, respectively,
\begin{eqnarray*}
&& V_x=\frac{\delta m^2}{2E}\sin 2\theta_0,\;V_y=0,\; V_z \simeq -\frac{\delta m^2}{2E}\cos 2\theta_0-\frac{14\pi^2}{45\sqrt{2}}\frac{G_F}{M^2_Z}E T^4_{\nu_a} + \frac{14\pi^2}{45\sqrt{2}}\frac{G_X}{M^2_X}E T^4_{\nu_s}, \nonumber \\
&& \Gamma_a \simeq G^2_F E T^4_{\nu_a},\;
 \Gamma_s \simeq G^2_X E T^4_{\nu_s},\;
  D\simeq \frac{1}{2}\left(\Gamma_a + \Gamma_s\right).
\end{eqnarray*}
From the QKEs, we can recover neutrino oscillation in vacuum if we take the non-interacting limits, $G_F\rightarrow 0,\;G_X\rightarrow 0$,
\begin{equation*}
\dot{P}_0=0,\;\;\dot{\vec{P}}=\vec{V}\times \vec{P},
\end{equation*}
which describe the precesses of $\vec{P}$ around $\vec{V}$. Non-zero $\Gamma_i$'s effects are to repopulate different momentum modes to reach thermal distribution and $D$-terms would damp and shrink $\vec{P}$. As we can see in the above equations, the introduced new self-interaction leads to $\Gamma_s\neq 0$ and contributes to $V_z$ and $D$ in the above QKEs. 

\begin{figure}[t]
\includegraphics[width=0.6\textwidth]{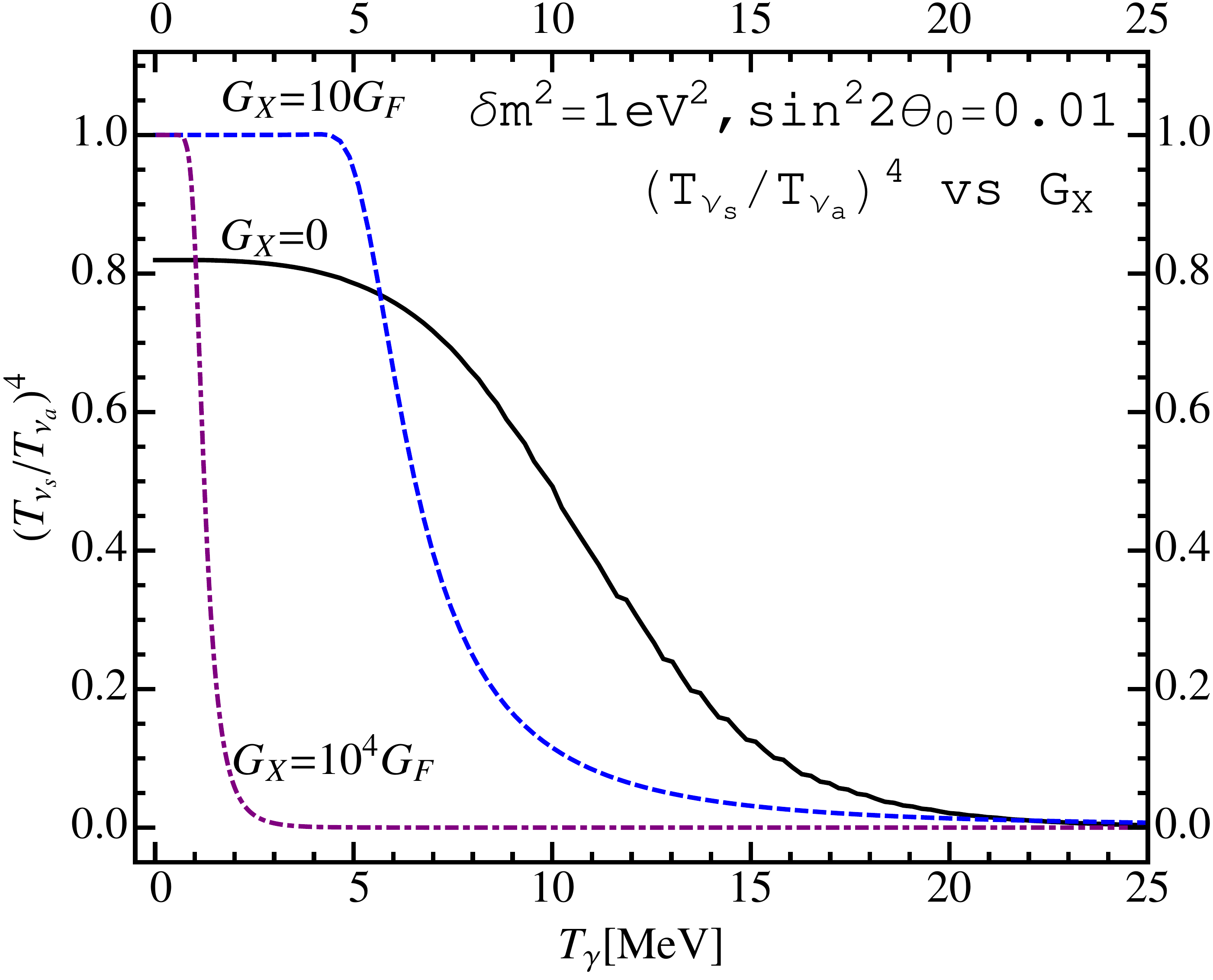}
\caption{Evolution of $\delta N_{\textrm{eff}}$ as temperature $T_\gamma $ decreases. $\delta N^{\textrm{bbn}}_{\textrm{eff}}$ depends on the self-interaction strength $G_X$. Black curve shows the non-interacting case, $G_X=0$. The self-interaction can suppress the production of sterile neutrino at high temperature, but lead to flavor equilibrium at later time. Increasing the strength of self-interacting would delay the equilibrium time.
\label{fig:neffGx}}
\end{figure}

The introduced self-interactions for sterile neutrinos have two effects. One is to block the thermalization at high temperature. The other effect is the collisions that lead a scattering-induced decoherent production at later time.
In Fig.~\ref{fig:neffGx}, we show the non-interacting case with a black curve and compare it with two interacting cases, $G_X=10G_F$ and $G_X=10^4G_F$. If $G_X$ is large, for instance $G_X=10^4G_F$, $\nu_s$ can only be partially thermalized with $T_{\nu_s}<T_{\nu_a}$ when $T_\gamma>2$MeV. However, if $G_X$ is not large enough, for instance $G_X=10G_F$, it will block the thermalization first at high temperature but enhance the production of sterile neutrino at a later time even before BBN epoch, $T_\gamma\simeq 5$MeV. The $G_X=10^4G_F$ case, however, has shown that the equilibrium time is later than the $G_X=10G_F$ case, less than $1\MeV$. Generally, increasing $G_X$ would delay the equilibrium time.

A simplified picture to understand these two effects is to use the effective mixing angle, 
\begin{equation}
\sin^{2}2\theta_{\mathrm{eff}}=\frac{\sin^{2}2\theta_{0}}{\left(\cos2\theta_{0} - \frac{2E}{\delta m^{2}}V_{\mathrm{eff}}\right)^{2}+\sin^{2}2\theta_{0}},
\end{equation}
where $E$ is the energy of oscillating neutrino, $\dfrac{\delta m^{2}}{2E} \cos{2\theta_0}$ is usually called the vacuum oscillation term~\cite{Akhmedov:1999uz}, and  $V_{\mathrm{eff}}=V^{\mathrm{SM}}_{\mathrm{m}}-V^{\textrm{NEW}}_{\mathrm{m}}$ is the matter potential. Simple analysis would give
\begin{equation}\label{eq:matterpotential}
V^{\mathrm{SM}}_{\mathrm{m}}\sim \frac{G_F}{M_Z^2} E T_{\nu_a}^4,\; V^{\textrm{NEW}}_{\mathrm{m}}\sim \frac{G_X}{M_X^2}E T_{\nu_s}^4,
\end{equation}
which highly depend on the temperature. A familiar case, MSW effect~\cite{Wolfenstein:1977ue,Smirnov:1986gs,Smirnov:1986wj}, happens when $V_{\mathrm{eff}} = \cos{2\theta_0}\delta m^{2}/2E$, leading to a maximal mixing angle $\theta_{\mathrm{eff}}=\pi/4$. However, when the matter potential is much larger than the vacuum term, the mixing angle is effectively suppressed, 
\begin{equation}
\sin^{2}2\theta_{\mathrm{eff}}\ll \sin^{2}2\theta_{0}, \; \textrm{when } 
\left|V_{\mathrm{eff}}\right| \gg \left|\dfrac{\delta m^{2}}{2E}\cos{2\theta_0}\right|,
\end{equation}
which can efficiently block the production of sterile neutrinos. 
The value of $T_{\nu_s}$ to block production at BBN time can be roughly estimated as follows:
\begin{equation*}
V_{\mathrm{eff}}\sim \frac{G_X}{M^2_X}E T^4_{\nu_s} >\frac{\delta m^2}{2E}\Rightarrow \frac{T_{\nu_s}}{\textrm{MeV}}>\left(\frac{\delta m^2}{2E^2}\frac{M^2_X}{G_X}\right)^{1/4}.
\end{equation*}
Take $G_X\sim 10^4 G_F$ and $M_X\sim 1$ GeV, we get $T_{\nu_s}\sim 10^{-3}$MeV around BBN time. This is what wee see the smallness of $T_{\nu_s}$ all the way to BBN time from the dot-dashed line in Fig.~\ref{fig:neffGx}.

As the Universe cools down, the matter potential $V_\textrm{eff}$ gets smaller very quickly. When $|V_{\mathrm{eff}}| < \left|\dfrac{\delta m^{2}}{2E}\cos{2\theta_0}\right|$, matter effect can be neglected and $\theta_{\textrm{eff}}\simeq \theta_0$.
Again, we discuss in the simplified framework of two-flavor case, $\nu_a$-$\nu_s$ or $\nu_1$-$\nu_2$ in the mass eigenstates,
\[
\nu_a=\cos \theta_{\mathrm{eff}} \ \nu_1  - \sin \theta_{\mathrm{eff}} \ \nu_2, \;
\nu_s=\sin \theta_{\mathrm{eff}} \ \nu_1 + \cos \theta_{\mathrm{eff}} \ \nu_2. 
\] 
Before BBN, because of the highly suppressed $\theta_{\mathrm{eff}}$, there are mostly $\nu_a$ or $\nu_1$ neutrinos in the thermal bath, and a small amount of $\nu_s$ or $\nu_2$ states. After $\theta_{\mathrm{eff}}$ is not suppressed any more, $\nu_1$ has the scattering process to produce $\nu_2$ through $\nu_1+\nu_2\rightarrow \nu_2 + \nu_2$ with rate $\Gamma \sim G^2_X T^5_{\nu_{s}}\sin ^2 2\theta_0$, and $\nu_1+\nu_1\rightarrow \nu_2 + \nu_2$ with rate $\Gamma \sim G^2_X T^5_{\nu_{a}}\sin ^4 2\theta_0$. If $\Gamma$ is larger than Hubble parameter $H=\sqrt{\dfrac{8\pi G}{3}\rho_R}$, then $\nu_1$ and $\nu_2$, or  $\nu_a$ and $\nu_s$, will reach equal numbers quickly. 
Since the sterile neutrinos' self-interaction will induce rapidly elastic scattering, $\nu_s \nu_s \rightarrow \nu_s \nu_s$, it can redistribute momenta among sterile neutrinos. As long as the scattering rate $\Gamma_X\sim G^2_X T_{\nu_s}^5$ is larger than the Hubble parameter $H$ ($G$ is the Newton's constant), sterile neutrinos will soon reach the Fermi-Dirac distribution, leading to the flavor equilibrium, $n_{\nu_s}=n_{\nu_a}$.  For $G_X=10^8 G_F$ and $M_X=1.2$MeV it was estimated that the equilibrium would be approached around $T_\gamma\sim 40$KeV~\cite{Mirizzi:2014ama}, although numerically being a formidable challenge for such a large interaction. 

\begin{figure}[t]
\includegraphics[width=0.45\textwidth,height=0.5\textwidth]{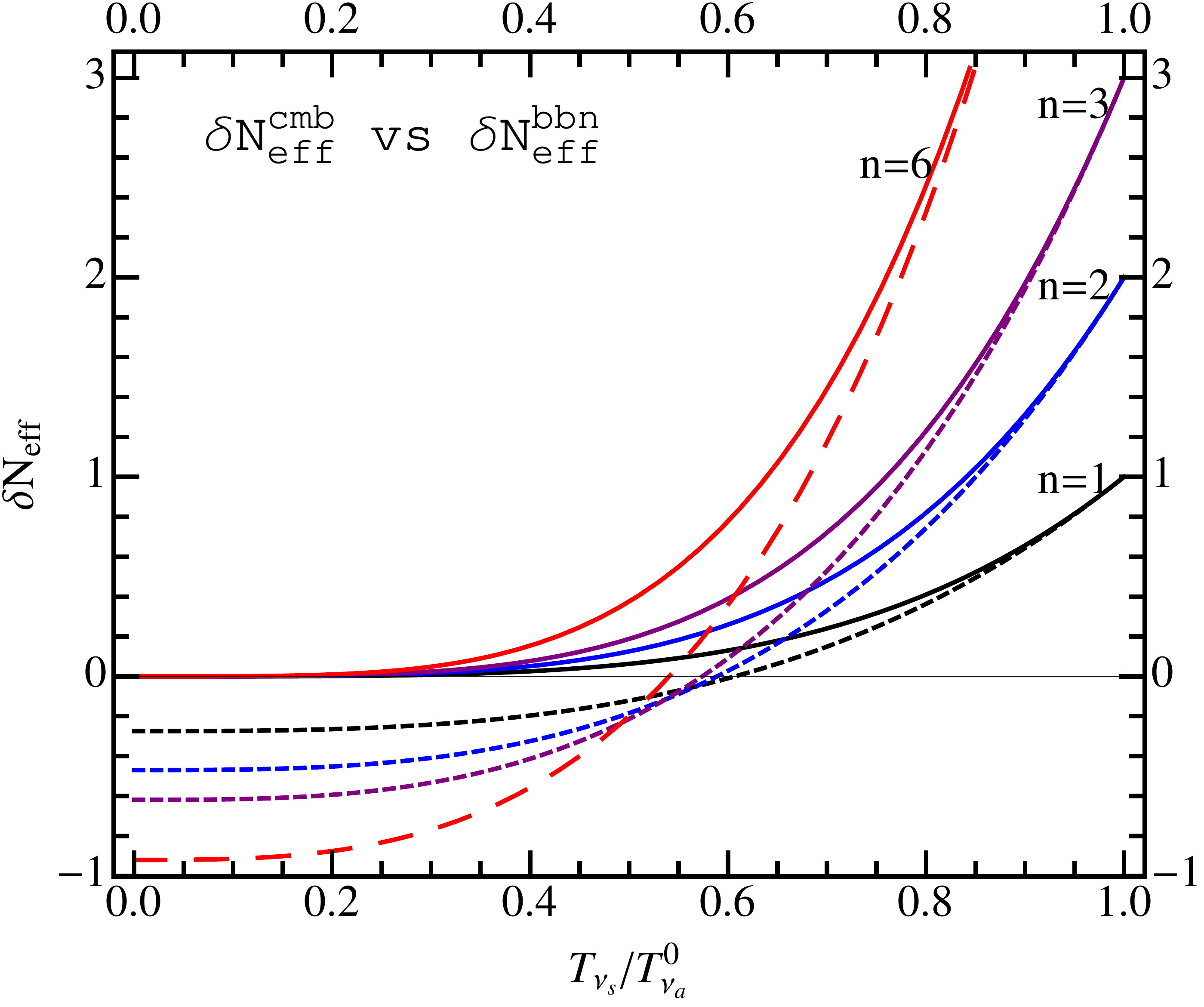}
\includegraphics[width=0.485\textwidth,height=0.502\textwidth]{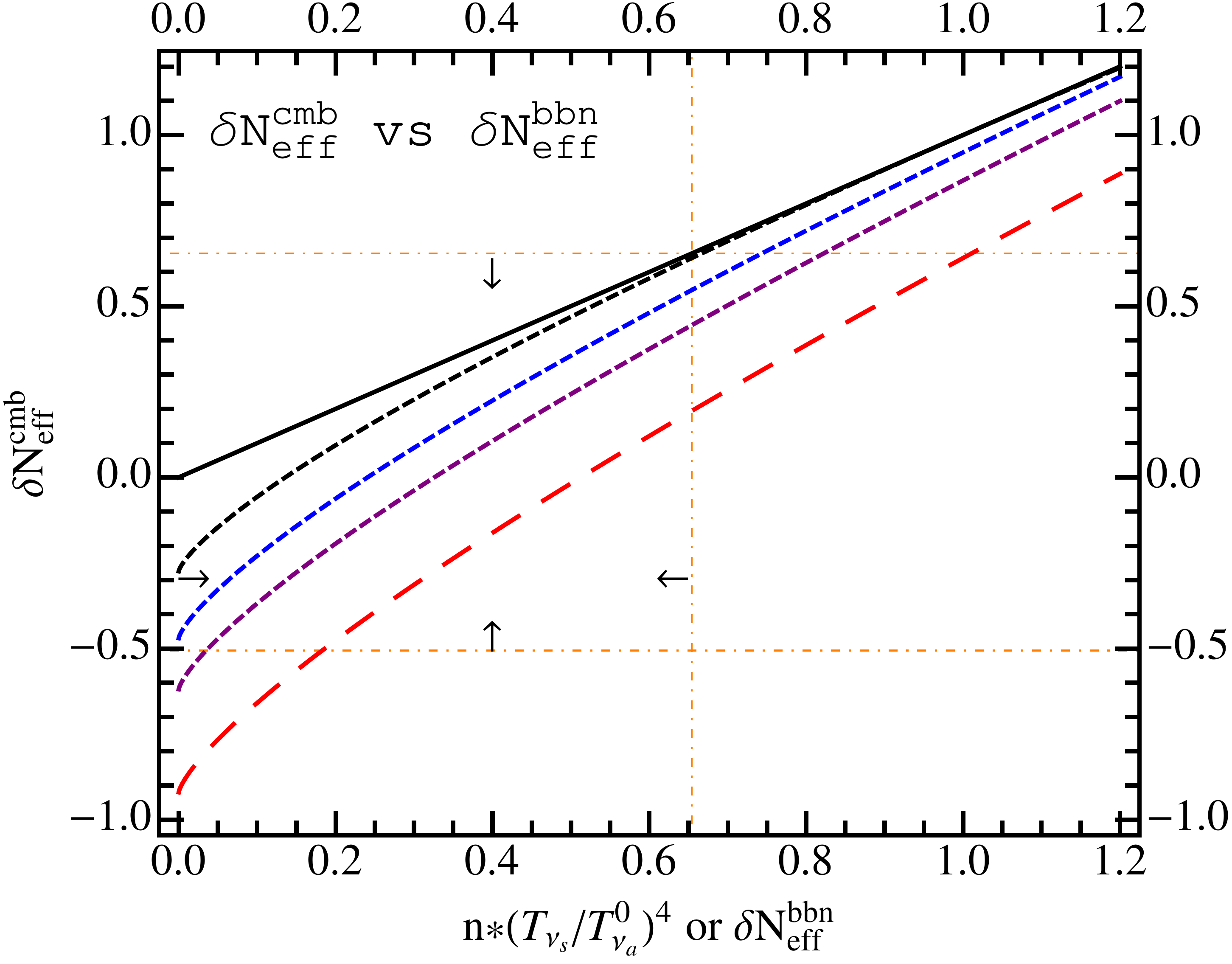}
\caption{$\delta N^{\textrm{bbn}}_{\textrm{eff}}$ vs $\delta N^{\textrm{cmb}}_{\textrm{eff}}$. (Left panel)We choose several cases for the number of sterile species, as indicated by $n$. For each case, the solid curve shows $\delta N^{\textrm{bbn}}_{\textrm{eff}}$ while the dashed one gives $\delta N^{\textrm{cmb}}_{\textrm{eff}}$. Sizable differences can arise in the low $T_{\nu_s}/T^0_{\nu_a}$ region. (Right panel)$\delta N^{\textrm{cmb}}_{\textrm{eff}}$ as function of $\delta N^{\textrm{bbn}}_{\textrm{eff}}$. The solid black line is for $\delta N^{\textrm{cmb}}_{\textrm{eff}}=\delta N^{\textrm{bbn}}_{\textrm{eff}}$ in non-interacting case, and from up to down, dashed lines correspond to $n=1,2,3,6$. Dot-dashed orange lines mark the current bounds from {\it Planck}~\cite{Planck:2015xua}.
\label{fig:dNeff}
}
\end{figure}

Before a detailed discussion on $N_{\textrm{eff}}$ at or after CMB time, we should note that the above investigation only took into account the sterile neutrinos produced from oscillations. However, oscillation is not the only contributing process. Generally, we expect there is a whole dark sector accompanying with self-interacting sterile neutrinos. Once the whole dark sector is decoupled from the SM thermal both, entropy is transfered to sterile neutrinos(and other particles in thermal equilibrium with sterile neutrinos). We may call this part as the ``primordial'' portion (denoted as $\delta N_{\textrm{eff}}^0$) which depends on the physical degrees of freedom in the dark sector and decoupling temperature, therefore model-dependent, see, for example, Refs.~\cite{Bringmann:2013vra, Ko:2014bka} for concrete models. Since at high temperature oscillation is effectively blocked by the large matter potential, the effects of these primordial sterile neutrinos are to change the initial condition for QKEs and lift up the curves at high temperature by $\delta N_{\textrm{eff}}^0$ in Fig.~\ref{fig:neffGx}. Hence, if not stated explicitly, we shall treat $T_{\nu_s}$ at BBN time as a free parameter in the rest of our discussion.

Now, we are in a position to discuss the effect on $N_{\textrm{eff}}$ at or after CMB time. Assume there are $n$ sterile species with common temperature $T_{\nu_s}$,
if sterile and active neutrinos reach the equilibrium when they are still relativistic, they would have the same temperature $T_{\nu}$ determined by the conservation of neutrino number or entropy density,
\begin{equation}\label{eq:entropy}
3\times \left(T_{\nu_a}^{0}\right)^3 + n\times T_{\nu_s}^3 = \left(3 + n\right)\times T_{\nu}^3,
\end{equation}
where $n$ is the number of sterile species that have self-interactions~\footnote{We note that n=1 case has been discussed in the latest version of Ref.~\cite{Mirizzi:2014ama} whose results agree with ours.}.

With the new temperature $T_{\nu}$, we can calculate $\delta N_{\textrm{eff}}$ at CMB time with Eq.~\ref{eq:deltaNeff}, 
\begin{equation}\label{eq:neffcmb}
\delta N_{\textrm{eff}}^{\textrm{cmb}}=\left(3+n\right)^{-1/3}\times \left[3+n\times \left(\dfrac{T_{\nu_s}}{T^0_{\nu_a}}\right)^3\right]^{4/3}-3,
\end{equation}
in comparison with
\begin{equation}\label{eq:neffbbn}
\delta N_{\textrm{eff}}^{\textrm{bbn}} = n\times\left(\dfrac{T_{\nu_s}}{T^0_{\nu_a}}\right)^4. 
\end{equation}

If sterile neutrinos were fully thermalized at BBN time, we would have $T_{\nu_s}=T^0_{\nu_a}$, and Eq.~\ref{eq:neffcmb} gives the same result as Eq.~\ref{eq:neffbbn} does. However, for partially thermalized $\nu_s$, $T_{\nu_s}< T^0_{\nu_a}$. As shown in Fig.~\ref{fig:dNeff}, it is evident that $\delta N^{\textrm{cmb}}_{\textrm{eff}}\leq \delta N^{\textrm{bbn}}_{\textrm{eff}}$, and that the difference can be significant in the low $T_{\nu_s}/T^0_{\nu_a}$ region and it increases as $n$ gets bigger. An interesting observation is that $\delta N_{\textrm{eff}}^{\textrm{cmb}}$ can even be negative for small values of  $T_{\nu_s}/T^0_{\nu_a}$. If future experiment data indicate a deficit of $N_{\textrm{eff}}$, it would be natural to consider the scenario that active neutrinos are mixing with self-interacting sterile species.

In the right panel of Fig.~\ref{fig:dNeff}, we show $\delta N^{\textrm{cmb}}_{\textrm{eff}}$ as functions of $\delta N^{\textrm{bbn}}_{\textrm{eff}}$ for different $n$. The solid black line is for $\delta N^{\textrm{cmb}}_{\textrm{eff}}=\delta N^{\textrm{bbn}}_{\textrm{eff}}$ in non-interacting case, and from up to down, dashed lines correspond to $n=1,2,3,6$, respectively. Dot-dashed orange lines mark the current $95\%$ CL preferred ranges for $N_{\textrm{eff}}$ from {\it Planck}~\cite{Planck:2015xua},
\begin{equation}
N_{\textrm{eff}}=3.11^{+ 0.59}_{-0.57} \;\;\;\; \textrm{He+{\it Planck} TT+lowP}.
\end{equation}
Since $\delta N^{\textrm{bbn}}_{\textrm{eff}}\geq 0$ in our scenario, the available parameter space is inside the region with arrows indicated.


\section{Cosmological Neutrino Mass Bounds}\label{sec:massbounds}

In this section, we show how the conflict between eV sterile neutrino and cosmological mass bounds can be relaxed when more than one light sterile species are introduced.

Cosmological bounds on the neutrino masses from the combination of CMB, large scale structure and distance measurements are constraining the following effective quantity~\cite{Planck:2013zuv},  
\begin{equation}\label{eq:effectiveMass}
m^{\textrm{eff}}_{\nu}\equiv \frac{\sum_i n_{\nu_i}m_{\nu_i}}{n^0_{\nu_a}}=\sum_{i}\left(\frac{T_{\nu_i}}{T_{\nu_a}^0}\right)^3 m_{\nu_i}
\simeq 94.1 \eV \times \Omega_{\nu}h^2,
\end{equation}
where $n_{\nu_i}$ stand for $\nu_i$'s number density, $n^0_{\nu_a}$ for the value of active neutrino in standard cosmology, $\Omega_{\nu}h^2$ accounts for its energy density fraction in the Universe. After the later flavor equilibrium discussed above, all neutrinos share the same temperature, $T_{\nu_i}=T_\nu$. Using the same minimal setup as we did in last section, we assume only one sterile state has eV-scale mass and all the others are almost massless. So we can reduce the above summation only over the heaviest one, $i=4$,
\begin{equation}
m^{\textrm{eff}}_{\nu}\simeq \left(\frac{T_{\nu}}{T_{\nu_a}^0}\right)^3 m_{\nu_4}.
\end{equation}

\begin{figure}[t]
\includegraphics[width=0.46\textwidth]{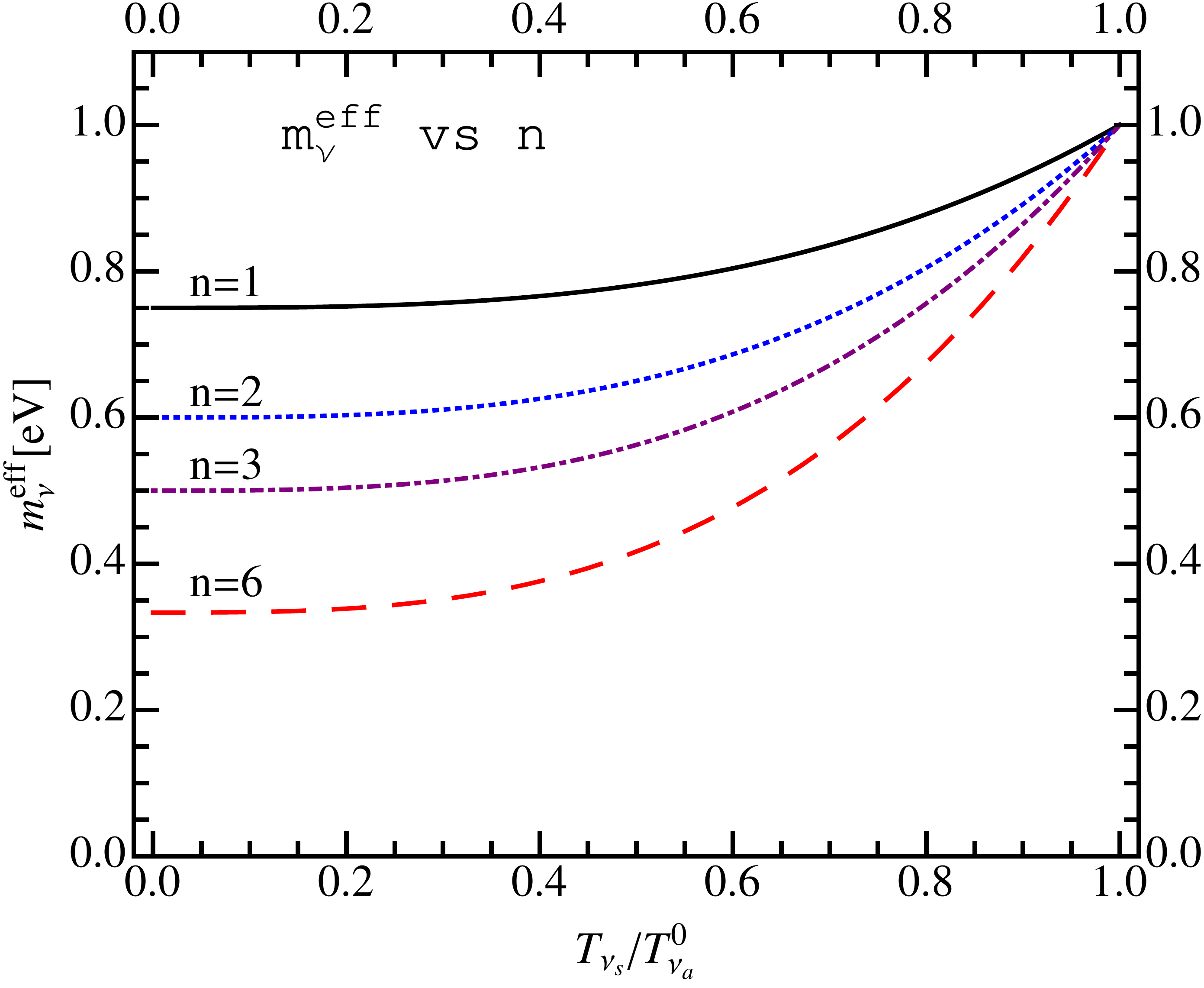}
\includegraphics[width=0.46\textwidth]{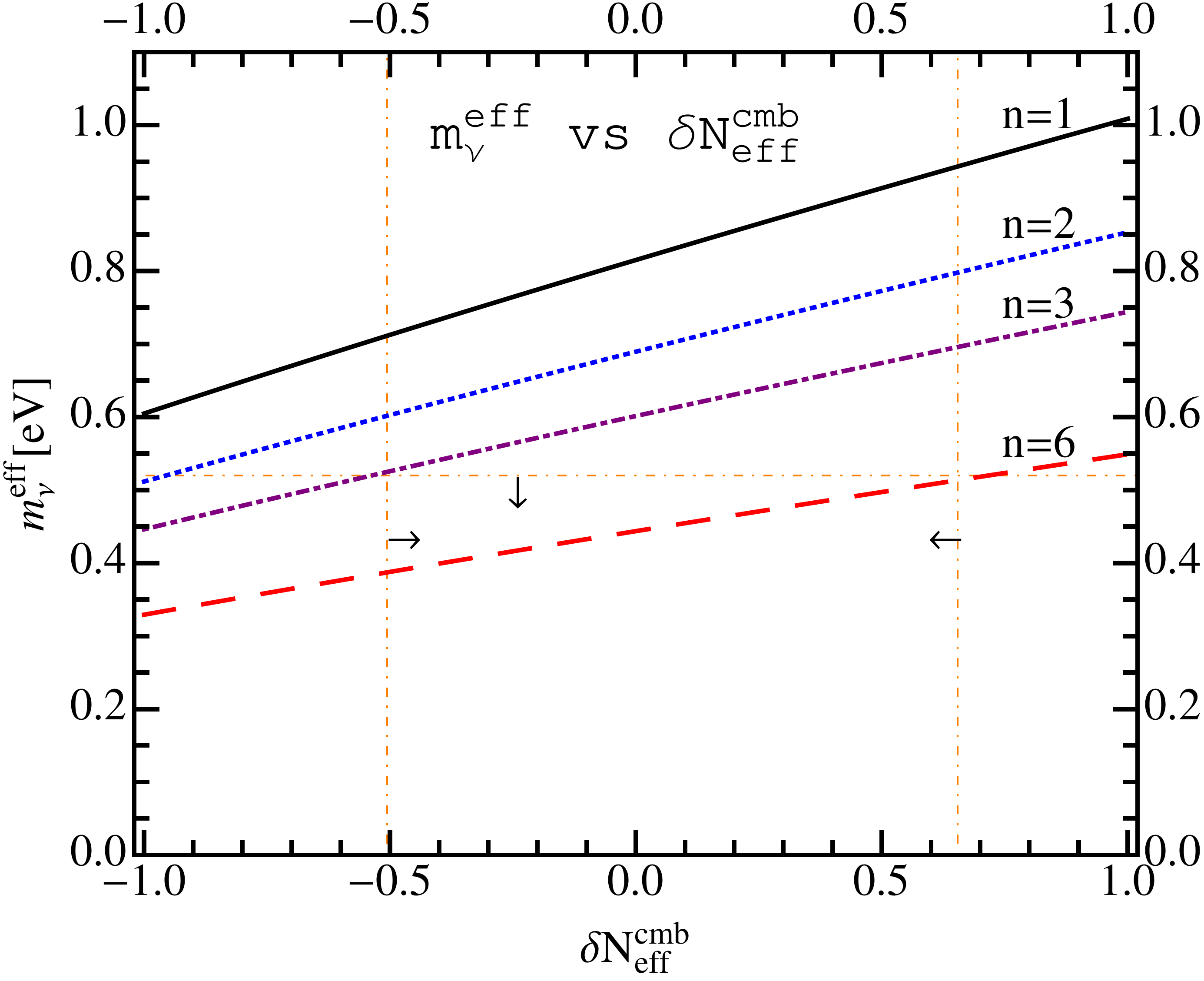}
\caption{(Left panel)$m^{\textrm{eff}}_{\nu}$ vs $n$. $m^{\textrm{eff}}_{\nu}$ is a model-dependent quantity. We show how it changes as $T_{\nu_s}/T^0_{\nu_a}$ varies in four cases with different $n$. $m_{\nu_4}=1$eV is assumed to be dominant on the mass. (Right panel)$m^{\textrm{eff}}_{\nu}$ as functions of $\delta N^{\textrm{cmb}}_{\mathrm{eff}}$ and $n$. Parameter space inside region marked with arrows is still allowed.  \label{fig:meff}}
\end{figure}

In Fig.~\ref{fig:meff}, we plot how $m^{\textrm{eff}}_{\nu}$ changes with $ T_{\nu_i}/T_{\nu_a}^0$ in four cases, $n=1,2,3,6$. When more light states are added with but fixed $\delta N_{\textrm{eff}}$, the individual number density of each species is decreased. Therefore, the total number of the heaviest state $\nu_4$ is reduced and $m^{\textrm{eff}}_{\nu}$ then gets smaller correspondingly. Other light states are just radiations and red-shifted, contributing only negligibly in late Universe. 

Now, we compare with the cosmological bounds. We should note that cosmological bounds on sterile neutrino mass and abundance depend on the cosmological models and the chosen data set~\cite{Hamann:2011ge,Boehm:2014ksa,Melchiorri:2014zza,Dvorkin:2014lea,Zhang:2014dxk,Hannestad:2014apa,Verde:2014sia,Giunti:2014dla,Cyr-Racine:2013jua}~\footnote{
Currently, there is a very loose constraint on self-interaction of active neutrinos from CMB data~\cite{Cyr-Racine:2013jua}, around $\sim 10^8 G_F$. The bound on self-interaction of sterile neutrino can be inferred and should be similar. }. Varying $N_{\textrm{eff}}^{\textrm{cmb}}$ and $ m^{\textrm{eff}}_{\nu}$ only, and using the $Planck$+WP+HighL data combination, the latest result from {\textit{Planck}} collaboration was able to give bounds with $95\%$ CL~\cite{Planck:2015xua}, 
\begin{equation}\label{eq:bounds2}
 2.53<N_{\textrm{eff}}^{\textrm{cmb}}<3.7,\; m^{\textrm{eff}}_{\nu}<0.52 \textrm{eV}.
\end{equation}
As we show in right panel of Fig.~\ref{fig:meff}, if we the face value of the above constraint~\footnote{Note that the above constraint is only intended for non-interacting sterile neutrinos. Self-interactions may change the ``free-streaming'' scale, see Ref.~\cite{Chu:2015ipa} for example. Since no analysis with self-interacting sterile neutrino is available, here we only took the face value from {\textit{Planck}}~\cite{Planck:2015xua}.}, then $n=3$ is on the intersect point with marginal status and is the minimal number of introduced sterile species. This amusing accidental agreement recovers the symmetry between active and sterile neutrinos. If future experimental pushed the upper limit further stringent, from the trend shown in the four cases of Fig.~\ref{fig:meff}, it is easy to introduce more light sterile states in the discussed scenario to relax the cosmological bounds.
When putting the lower bounds on $N_{\textrm{eff}}$, we should be aware of the assumption that no other relativistic particles contribute as radiations. In cases where there are quite a mount of massless particles such as Goldstone or Majaron particles, the lower bounds on $N_{\textrm{eff}}$ then do not apply and $n<3$ will be allowed.

\section{Conclusion}
In this paper, we have discussed a scenario that eV sterile neutrinos are partially thermalized before BBN era but equilibrated with active ones in later time. A mechanism to realize such a scenario is to introduce secret self-interactions for sterile neutrinos. The self-interactions can induce large matter potentials at high temperature, suppress the mixing angle and block the production of sterile neutrinos from oscillations. They can also lead to a rapid scattering-induced decoherent production of sterile neutrinos at later times before CMB. When flavor equilibrium between active and sterile species is approached, it surprisingly leads to a decrease of $N_\textrm{eff}$. 

We also discussed how the conflict with cosmological neutrino mass bounds can be relaxed in this scenario. We have found that, the more light sterile species we add, the less constrained they would be. If we take the latest Planck bounds~\cite{Planck:2015xua},
$ N_{\textrm{eff}}^{\textrm{cmb}}<3.7$ and $m^{\textrm{eff}}_{\nu}<0.52 \textrm{eV}$, at least three sterile species are needed to evade such a constraint.

\begin{acknowledgments} 
The author is grateful to Rasmus Hansen for helpful discussions and sharing his code. This work is supported in part by National Research Foundation of Korea (NRF) Research Grant 2012R1A2A1A01006053.
\end{acknowledgments}


\providecommand{\href}[2]{#2}\begingroup\raggedright\endgroup

\end{document}